\documentclass[letterpaper,aps,twocolumn,showpacs,superscriptaddress,floats,prl]{revtex4}

\usepackage[english]{babel}
\usepackage[latin1]{inputenc}
\usepackage{amsmath}
\usepackage{amssymb}
\usepackage{dsfont}
\usepackage{epsfig}
\usepackage{natbib}
\usepackage{graphicx}
\usepackage{bm}
\usepackage{color}

\begin{document}

\title{Quantum critical elasticity}

\author{Mario Zacharias}
\affiliation{Institut f\"ur Theoretische Physik, Universit\"at zu K\"oln,
Z\"ulpicher Str. 77, 50937 K\"oln, Germany
}
\author{Indranil Paul}
\affiliation{Laboratoire Mat\'eriaux et Ph\'enom\`enes Quantiques,
Universit\'e Paris Diderot-Paris 7 \& CNRS, UMR 7162, 75205 Paris, France
}
\author{Markus Garst}
\affiliation{Institut f\"ur Theoretische Physik, Universit\"at zu K\"oln,
Z\"ulpicher Str. 77, 50937 K\"oln, Germany
}

\date{\today}

\begin{abstract}
We discuss elastic instabilities of the atomic crystal lattice at zero temperature.
Due to long-range shear forces of the solid, at such transitions the phonon velocities vanish, if at all, only along certain crystallographic directions, and, consequently, the critical phonon fluctuations are suppressed to a lower dimensional manifold and governed by a Gaussian fixed-point. In case of symmetry-breaking elastic transitions, a characteristic critical phonon thermodynamics arises that is found, e.g., to violate Debye's $T^3$-law for the specific heat.
We point out that quantum critical elasticity is triggered whenever a critical soft mode couples linearly to the strain tensor.
In particular, this is relevant for the electronic Ising-nematic quantum phase transition in a tetragonal crystal as discussed in the context of certain cuprates, ruthenates and iron-based superconductors.
%
\end{abstract}

\date{\today}

\maketitle


Quantum fluctuations close to an instability of the ground state result in fascinating, exotic behavior
even at finite $T$ that is often at odds with conventional properties of materials~\cite{SachdevKeimer-2011}.
However, most of the recent interest focused on instabilities associated with the electronic degrees of freedom
in metals and insulators~\cite{Reviews-Metals,Sachdev-2008,Giamarchi-2008}. In contrast, zero temperature instabilities of the atomic crystal lattice have attracted much less attention. While distortive instabilities associated with the softening of an optical phonon have been studied in the context of quantum
critical paraelectrics \cite{paraelectric,Cano2004,Rowley2014}, quantum phase transitions involving the elastic
degrees of freedom, namely the uniform strains and the acoustic phonons, have remained largely unexplored
even though such transitions are ubiquitous in various phases of matter such as insulators, metals and superconductors. It is the aim of this work to fill this gap, and to study the
critical thermodynamics associated with elastic {\it quantum} criticality (EQC).

Elastic {\it classical} criticality, i.e., elastic instabilities at a finite critical temperature $T_c$ have been already studied and classified by Cowley \cite{Cowley:1976} and Schwabl and collaborators \cite{Folk:1976,Folk79,Schwabl:1980,Schwabl:1996}. These works identified the importance of shear rigidity, a property that distinguishes crystals from liquids and gases. This rigidity and the concomitant long-range forces restrict criticality to an $m$-dimensional subspace in a $d$-dimensional Brillouin zone, with $m \leq d$, making the transitions mean-field type for $d=3$ \cite{Folk:1976,Chalker1980}. This physics is also crucial for EQC, setting them apart from the paradigm of conventional quantum phase transition, i.e., without any long range forces, involving electrons in metals and insulators \cite{SachdevBook}. Compared to finite $T$ elastic transitions, in EQC one needs to take into account the dynamics of the acoustic phonons. As a result, we find, for example, that Debye's $T^3$-law for the specific heat $C_p = T (\partial S/\partial T)_p$ is violated in a characteristic manner close to symmetry-breaking EQC.

The following are our main results. From a study of the statics and the dynamics of the critical acoustic phonons, we construct the scaling form of the free energy associated with all the different universality classes of EQC. This allows to obtain, besides the specific heat, other thermodynamic quantities such as the thermal expansion $\alpha = (1/V)(\partial V/\partial T)_p$, and the ratio $\Gamma = \alpha/C_p$. The latter is a variant of the well-known Gr\"uneisen parameter \cite{Grueneisen}, and it can be identified with the relative change of temperature upon adiabatically changing the pressure, $\Gamma = 1/(V_m T) (\partial T/\partial p)_S$ with the molar volume $V_m$. The quantity $\Gamma$ has proven useful in the investigation of quantum criticality in general as it necessarily diverges at a pressure-tuned quantum critical point with characteristic power-laws~\cite{Zhu03,Garst:2005,Kuechler:2003,Gegenwart:2010,Loehneysen:2007,Gegenwart:2008}. Note that until now $\Gamma$ has been theoretically studied only in situations where the crystal lattice acts as a non-intrusive probe and itself remains non-critical. We also derive a general expression for the exponent $\theta$ characterizing the dependence of the transition temperature $T_c(r) \sim |r|^{\theta}$ on the tuning parameter $r$ of the quantum phase transition.

According to elasticity theory \cite{Landau}, the macroscopic stability of the crystal requires that all
eigenvalues of the elastic constant matrix $C_{ijkl}$ be positive.
This guarantees that the acoustic phonon velocities, determined by the eigenvalues of the dynamical matrix
$M_{ik}({\bf q}) = C_{ijkl} q_j q_l$ are finite, where ${\bf q}$ is the phonon momentum. At an elastic transition a specific eigenvalue of $C_{ijkl}$ vanishes and, depending on the degeneracy of this particular eigenvalue, the strain
order parameter is either a singlet, doublet or triplet of the irreducible representations of the crystal
class \cite{Cowley:1976,Folk:1976,Folk79,Schwabl:1980,Schwabl:1996}.
Importantly, at the instability the phonon velocity goes to zero only for momenta along certain high symmetry directions for which the phonon triggers only the critical strain mode.
However, for a generic direction the phonon excites the non-critical strains with finite elastic constants as well, and consequently their velocity stays finite at the transition.
Thus, elastic criticality can be classified as type 0, I or II depending on the
dimensionality $m=0,1,2$, respectively, of the critical phonon subspace \cite{Cowley:1976}. The fact that the phonons
remain non-critical in a $(d-m)$-dimensional subspace distinguishes elastic transitions from conventional ones, for which $m=d$. In the following we study two types of EQC, namely those that
involve breaking of a point group symmetry of the unit cell (in which case $m=1,2$), and those that do not (in which case $m=0$).

\begin{table}[t]
\centering
\begin{tabular}{|c|c|c|c|}
\hline
elastic transition & constant & strain & type
\\
\hline
orthorhombic $\to$ monoclinic & $c_{44}$ & $\varepsilon_{23}$ & I
\\
orthorhombic $\to$ monoclinic & $c_{55}$ & $\varepsilon_{13}$ & I
\\
orthorhombic $\to$ monoclinic & $c_{66}$ & $\varepsilon_{12}$ & I
\\
tetragonal $\to$ orthorhombic & $c_{11}-c_{12}$ & $\varepsilon_{11}-\varepsilon_{22}$ & I
\\
tetragonal $\to$ orthorhombic & $c_{66}$ & $\varepsilon_{12}$ & I
\\
tetragonal $\to$ mono- or triclinic & $c_{44}$ & $(\varepsilon_{23}, \varepsilon_{13})$ & I+II
\\
hexagonal $\to$  mono- or triclinic & $c_{44}$ & $(\varepsilon_{23}, \varepsilon_{13})$ & I+II
\\
\hline
\end{tabular}
\caption{Continuous symmetry-breaking elastic transitions \cite{Cowley:1976,Folk:1976,Schwabl:1980}. Second column: component
of the elastic constant matrix in Voigt notation that goes to zero at the transition; third column: the strain
order parameter; fourth column: type of the transition in the classification of Cowley \cite{Cowley:1976}.
Modifications arise for tetragonal crystals with a finite $c_{16}$.
}
\label{table1}
\end{table}

{\em
Symmetry-breaking elastic transitions --}
If the EQC can be associated with the breaking of a crystal symmetry, the strain order parameter assumes a zero expectation value in the symmetric, undistorted phase.  Depending on the presence or absence
of a cubic invariant in the Landau potential for the order parameter, the transition is expected to be of first
or second order. Of particular interest are second-order transitions that are accompanied by critical fluctuations which induce unusual behavior at finite $T$. In Table~\ref{table1} we list the elastic transitions associated with spontaneous crystal-symmetry breaking. Most of these transitions are of type I with a strain order parameter that is  a singlet. The exceptions are listed in the last two rows that possess a doublet order parameter and are characterized by phonon velocities that vanish within one- as well as two-dimensional subspaces, i.e., type I and II, respectively.

The Landau potential for the order parameter $\varepsilon$, being either a singlet or a doublet, reads
\begin{align} \label{MeanField}
\mathcal{V}(\varepsilon) = \frac{r}{2} \varepsilon^2 + \frac{u}{4!} (\varepsilon^2)^2 + \sigma \varepsilon.
\end{align}
A second-order quantum phase transition obtains for a positive quartic coupling $u>0$ if the tuning parameter $r$ goes
to zero, $r \to 0$, at $T=0$ and $\sigma=0$. This occurs when the corresponding elastic constant listed in the second
column of Table~\ref{table1} vanishes. The strain in general couples to an externally applied stress $\sigma_{ij}$.
However, in most of the cases the appropriate $\sigma$ is a shear stress.
The tuning parameter, $r=r(p)$, will in general depend on hydrostatic pressure $p$, which arises from anharmonicities that mix the irreducible representations,
in particular, from a third order term that couples
the order parameter $\varepsilon$ to the trace of the strain,
$\mathcal{V}_{\rm int} \sim$ tr$\{\varepsilon_{ij}\} \varepsilon^2$.

Decomposing the phonon wavevector, $\bf q = (\bf p, \bf k)$, into an $m$-dimensional
soft component $\bf p$ and
a non-critical $(d-m)$-dimensional component ${\bf k}$ with $m=1,2$ for type I and II, respectively, the phonon dispersion close to criticality, $r\to0^+$, assumes the anisotropic form \cite{Folk79}
\begin{align}
\omega^2 \sim r {\bf p}^2 + a {\bf p}^4 + b {\bf k}^2 + \dots
\end{align}
with finite constants $a$ and $b$, and the dots represent other terms not relevant for the following discussion.
In order to deal with this anisotropic spectrum, a possibility is to perform the substitution ${\bf k}^2 \to {\bf k}'^4$.
It amounts to introducing an effective spatial dimensionality $d_{\rm eff} = m +2 (d-m) = 2 d - m$ with $d=3$.
The resulting scaling $r \sim {\bf p}^2$ and $\omega^2 \sim {\bf p}^4, {\bf k}'^4$  determines the correlation
length exponent $\nu = 1/2$ and $z=2$, respectively \cite{Remark}.
As a result of the enhanced effective dimensionality $d_{\rm eff}$, the EQC is above its upper critical dimension
$d_{\rm eff} + z  = 8-m > d_c^+$ with $d_c^+ = 4$ for any $m=1,2$ and
is governed by the Gaussian fixed-point, thereby justifying the above scaling. Evaluating the Gaussian fluctuations of the critical phonon modes, we find that the resulting free energy can be cast in the scaling form
%
%
\begin{align} \label{Fcr}
\mathcal{F}_{\rm cr} = T^{\frac{d_{\rm eff} +z}{z}} f\Big(\frac{r}{T^{1/(\nu z)}}\Big),
\end{align}
where the function $f$ possesses the asymptotics $f(x) = const.$ for $x \to 0$ and
$f(x) \sim x^{\nu d_{\rm eff} - \nu z d} = x^{-m/2}$ for $x\to \infty$.

\begin{figure}
\begin{minipage}{.45\columnwidth}
\includegraphics[width=\columnwidth]{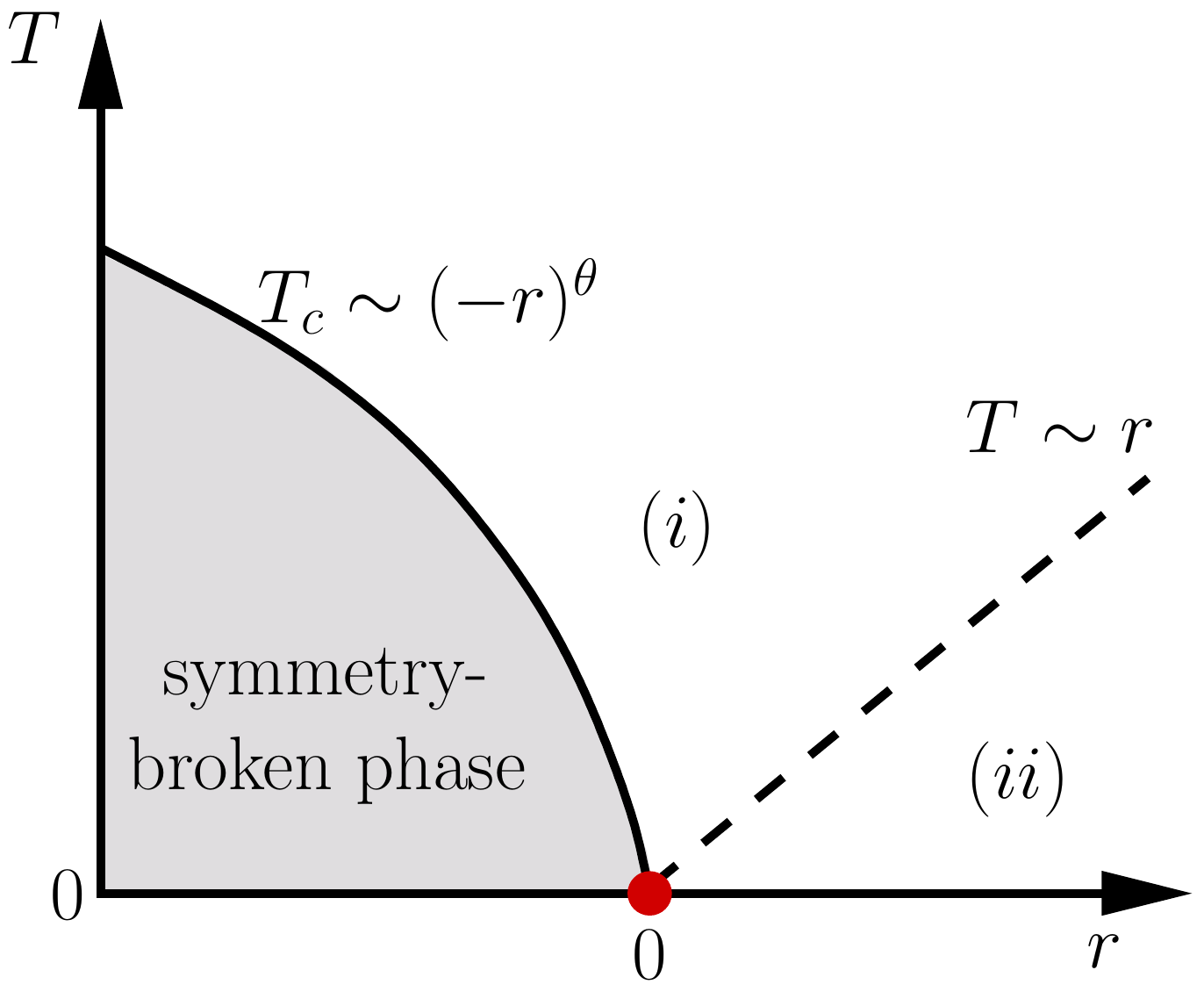}
\end{minipage}
\hspace{0em}
\begin{tabular}[c]{|c|c|c|}
\hline
 & $(i)$ & $(ii)$
\\
\hline
$C_{\rm cr} \sim$ & $T^{3-m/2}$ & $T^3 r^{-m/2}$
\\
$\alpha_{\rm cr} \sim$ & $T^{2-m/2}$ &  $T^3 r^{-m/2-1}$
\\
$\alpha_{\rm cr}/C_{\rm cr}$ & $\sim 1/T$ & $= \frac{m}{6} \frac{1}{V_m(p-p_c)}$
\\
\hline
\end{tabular}
\caption{{\em Left panel:} Phase diagram for a symmetry-breaking elastic quantum phase transition. The tuning parameter $r$ vanishes for tuning the corresponding elastic constant to zero, see second column of Table \ref{table1}. The critical phonon thermodynamics exhibits a crossover at $T \sim r$ giving rise to two regimes $(i)$ and $(ii)$. The phase boundary, $T_c \sim (-r)^\theta$, is determined by the exponent $\theta$, see text.
{\em Right panel:}
Critical phonon specific heat $C_{\rm cr}$, phonon thermal expansion $\alpha_{\rm cr}$ and Gr\"uneisen ratio
in the regimes $(i)$ and $(ii)$ for pressure tuning $r(T=0) \propto p-p_c$; $m=1,2$ for type I and II, respectively.
}
  \label{fig1}
\end{figure}

With the help of Eq.~\eqref{Fcr} the critical phonon thermodynamics is easily derived and summarized in Fig.~\ref{fig1}.
 In the quantum critical regime $(i)$ in Fig.~\ref{fig1}, we find, in particular, a critical contribution to the phonon specific heat, $C_{\rm cr} \sim T^{3-m/2}$, i.e., $C_{\rm cr} \sim T^{5/2}$
and $C_{\rm cr} \sim T^{2}$ for type I and type II transitions, respectively, signalling a breakdown of Debye's $T^3$-law.
The volume thermal expansion, $\alpha$, is determined by the pressure dependence of the tuning parameter $r$ so that $\alpha_{\rm cr} \sim T^{2-m/2}$
at $r=0$. The critical Gr\"uneisen ratio defined as $\Gamma_{\rm cr} = \alpha_{\rm cr}/C_{\rm cr}$ obeys
$\Gamma_{\rm cr}  \sim 1/T^{1/(\nu z)}$ with $\nu z = 1$ as expected from scaling considerations \cite{Zhu03}. In the limit $T \ll r$ of regime $(ii)$ in Fig.~\ref{fig1}, on the other hand,
Debye's $T^3$-law is recovered, however, with a critically enhanced prefactor, $C_{\rm cr} \sim r^{-m/2} T^3$. The Gr\"uneisen ratio diverges $\Gamma_{\rm cr} = \frac{m}{6} \frac{1}{V_m(p-p_c)}$ with a universal prefactor $m/6$ i.e., $1/6$
and $1/3$ for type I and II, respectively, where we used
$r(T=0,p) \propto p-p_c$ with the critical pressure $p_c$.
Note that the critical phonon signatures vanish with a relatively high power of $T$ and, in fact, might be subleading, for example, compared to gapless particle-hole excitations in metals.

We now turn to the discussion of the exponent governing the phase boundary $T_c \sim (-r)^\theta$ near the quantum critical
point in Fig.~\ref{fig1}. It is determined by the perturbative renormalization of the tuning parameter, $r \to R(T) = r + \delta r(T)$, that induces a $T$-dependence, and $R(T_c) = 0$.
For metals one generically expects a temperature dependence $\delta r \sim T^2$ so that $T_c \sim \sqrt{-r}$ and $\theta = 1/2$.
The situation is more interesting for solids where the phonons provide the leading contribution to thermodynamics,
e.g.~insulators or fully gapped superconductors.
In this case, the renormalization of $r$ is determined by the
self-interaction of critical phonon degrees of freedom. An explicit calculation, presented in the supplementary information (SI), yields
%
$\theta = \frac{2}{6-m}$. The $m=1$ result agrees with a previous calculation
in the context of quantum critical piezoelectric ferroelectrics \cite{Cano2004}.

\emph{Tetragonal-to-orthorhombic transition --}
As an illustration of a particular example, we discuss in some further detail the tetragonal-to-orthorhombic transition which is described by the order parameter $\varepsilon_{11} - \varepsilon_{22}$, see Table~\ref{table1}.
In this case, the critical phonon becomes soft for momenta along diagonals
in the $q_1$-$q_2$ plane, i.e., ${\bf q} \propto (1,\pm1,0)$, and the critical manifold is thus one-dimensional corresponding to
type I with $m=1$. For finite but small deviations from the $q_1 = q_2$ manifold the critical phonon dispersion is
given by, see SI,
\begin{align}
&\rho \omega^2({\bf q}) \approx
\\\nonumber
&\frac{c_{11} - c_{12}}{2} q_+^2 + \frac{2 c_{11} + 2 c_{12} + c_{66}}{4}  q_-^2
+ \frac{c_{44}}{4} q_3^2 + a q_+^4.
\end{align}
Here $q_{\pm} = (q_1 \pm q_2)/\sqrt{2}$, and $|q_+| \gg (|q_3|, |q_-|)$, and $\rho$ is the ionic mass density.
As $c_{11} - c_{12} \to 0$ at the transition, the dependence on $q_+$ is determined by the
last term $a q_+^4$ that derives from higher order terms of the strain potential.
In the vicinity of the second critical manifold $q_1 = - q_2$ the dispersion is obtained by interchanging $q_+ \leftrightarrow q_-$
in the above.
Importantly, the dispersion in other directions does not soften
and remains non-critical as
the remaining elastic constants stay finite.

An interesting aspect of this particular transition is that the  tetragonal symmetry
can also be explicitly broken by $\sigma = (p_2 - p_1)/2$, where $p_i$ is the uniaxial pressure along the $i=1,2$ direction.
This is reflected in the linear thermal expansion $\beta_i = - (1/V_m)(\partial S/\partial p_i)_T$. Whereas the
sum $\beta_1 + \beta_2$ is expected to show similar behavior as the volume thermal expansion, $\alpha$, the
uniaxial thermal expansion, defined by the difference
$\beta_\sigma \equiv \beta_2 - \beta_1 = - (1/V_m)(\partial S/\partial \sigma)_T$, is more singular.
Minimization
with respect to the order parameter yields $\varepsilon = - \sigma/r$ in the linear regime of small
$|\sigma| \ll \sqrt{r^3/u}$, and
$\varepsilon \sim \sigma^{1/3}$ in the non-linear regime of large $|\sigma| \gg \sqrt{r^3/u}$.
Taking the renormalization $r \to R(T)$ into account, one obtains in the former case $\beta^{\rm cr}_\sigma \sim \sigma \partial_T (1/R(T))$.
The resulting $T$-dependence is singular in regime $(i)$ with $\beta^{\rm cr}_\sigma \sim \sigma T^{-(1 + 1/\theta)}$,
where the exponent $\theta$ was introduced earlier, while in regime $(ii)$ $\beta_\sigma$ is analytic in $T$.
In the non-linear regime of $\sigma$ and small temperatures, on the other hand, the
effective modulus is determined by $r_{\rm eff} = \partial^2_\varepsilon \mathcal{V} \approx
\frac{u}{2}\varepsilon^2 \sim \sigma^{2/3}$ and
$\beta^{\rm cr}_\sigma \sim T^{3}
(\partial (r^{-1/2}_{\rm eff})/\partial \sigma) \sim T^3 \sigma^{-4/3}$.
The accompanying uniaxial Gr\"uneisen ratio at $r=0$ diverges $\beta^{\rm cr}_\sigma/C_{\rm cr} = 1/(9 V_m \sigma)$
with the universal prefactor $1/9$ as $\sigma$ is reduced.

{\em
Isostructural elastic transitions --}
The remaining elastic transitions not listed in Table~\ref{table1} are generically not of second order. Exceptions
are specific points in the phase diagram where the symmetry is enhanced by additional fine-tuning, and particular
interesting examples of this class are isostructural transitions. Here, the expectation value of a certain singlet
representation, $\varepsilon$, of the strain tensor which is itself invariant under all crystal symmetry operations,
changes in a critical manner.
An example is the isostructural volume collapse transition in a cubic crystal where the singlet, $\varepsilon = {\rm tr}\{\varepsilon_{ij}\}$, represents fluctuations of the volume, $\int_V d{\bf r}\varepsilon({\bf r}) = \delta V$.

\begin{figure}
\centering
\includegraphics[width=.4\columnwidth]{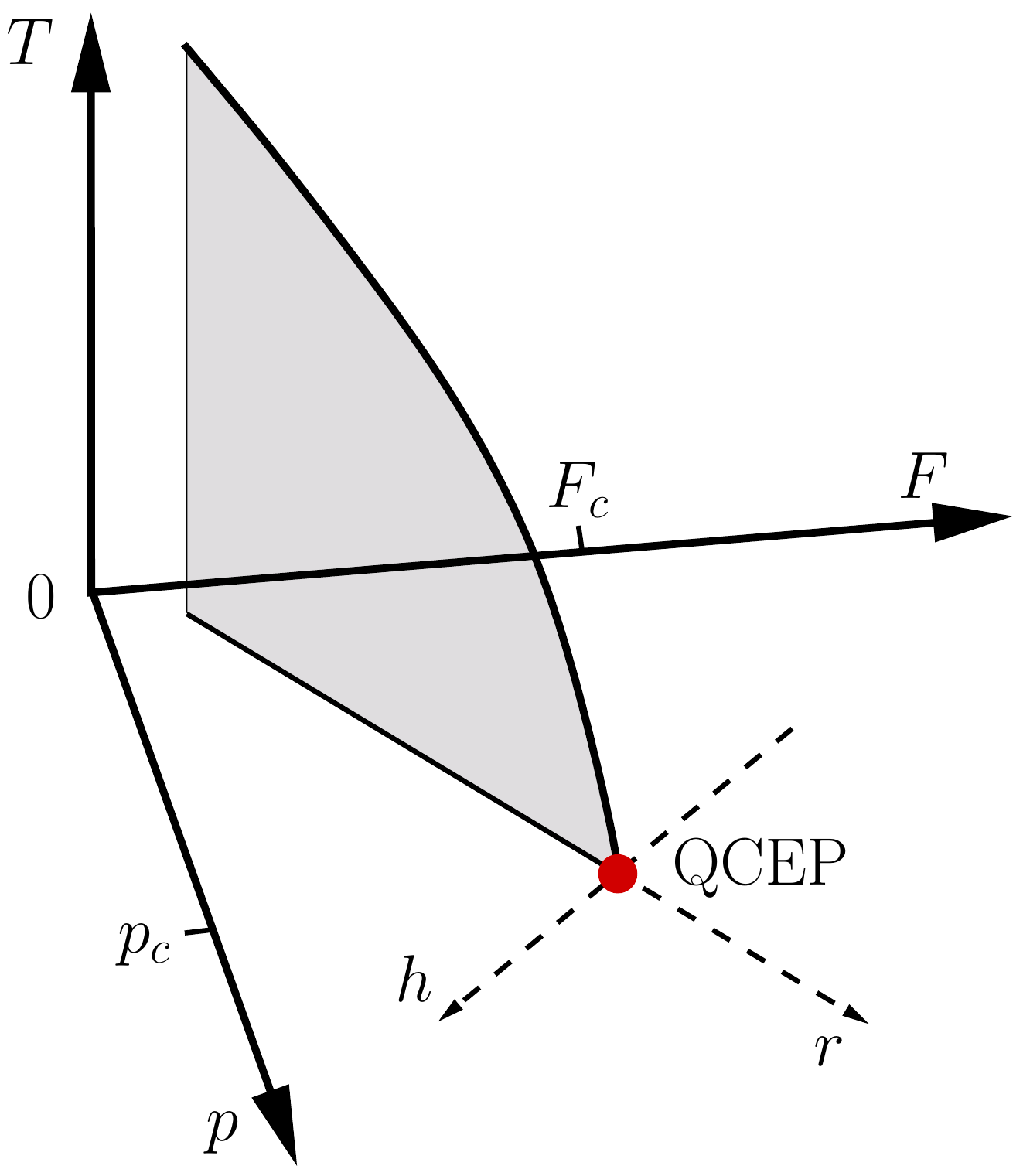}
\caption{Phase diagram with a solid-solid QCEP. A line of first-order solid-solid quantum phase transitions and a line of
finite temperature, $T$, second-order transitions meet at the QCEP and enclose a surface of first-order transitions at
finite $T$ (shaded area).
}
\label{fig2}
\end{figure}

The corresponding Landau potential generally contains all powers of $\varepsilon$. The cubic term, however,
can be made to vanish by appropriately shifting $\varepsilon \to \varepsilon + \varepsilon_0$ by a constant
$\varepsilon_0$ so that the potential assumes the same form as that of Eq.~\eqref{MeanField},
$\mathcal{V}(\varepsilon) = \frac{r}{2} \varepsilon^2 + \frac{u}{4!} \varepsilon^4 - h \varepsilon$,
%
where $h$ is to be identified, though, with an additional tuning parameter. In order to reach the second-order
quantum critical point both parameters, $h$ and $r$, must then be tuned to zero at $T=0$, for example, as a function
of an external field $F$ and pressure $p$. The criterion $h(F,p) = 0$ and $r(F,p) < 0$ defines a line of first-order
quantum phase transitions in the ($F,p$) phase diagram between isostructural solids characterized by different
expectation values of $\varepsilon$. This line terminates in a second-order quantum critical endpoint (QCEP) at a critical field, $F_c$,
and pressure, $p_c$, with $h(F_c,p_c) = r(F_c,p_c) = 0$, see Fig.~\ref{fig2}.

At this {\it solid-solid quantum critical point} a true mean-field transition occurs without critical microscopic
fluctuations. Due to the high symmetry of the order parameter $\varepsilon$, the isostructural transitions are
all of type 0
in the Cowley classification with the entire phonon sector being non-critical.
This peculiar aspect is rooted in the presence of
shear moduli that, in particular, distinguishes the solid-solid QCEP from the liquid-gas analogue.

A hallmark of solid-solid endpoints is the breakdown of Hooke's law. Minimizing
the potential
at $r = 0$ one
obtains $\varepsilon \sim h^{1/3} \propto (p - p_c)^{1/3}$ at the critical field $h(F_c,p) \propto p-p_c$ resulting
in a non-linear strain-stress relation with mean-field exponent $1/\delta = 1/3$ and a divergent compressibility
$\partial_p \varepsilon \sim |p - p_c|^{-2/3}$.  The resulting energy depends non-analytically on the tuning parameter
$h$, $\mathcal{V}_{\rm min} \sim |h|^{4/3}$.
However, due to the absence of critical microscopic fluctuations there is no diverging correlation length, and, as a
consequence, the usual scaling hypothesis for critical phenomena is not applicable. As a result, the thermodynamics at
finite $T$ for a solid-solid QCEP is non-universal and depends on the $T$-dependence of the tuning parameters,
e.g., $h = h(p,F,T)$, induced by non-critical degrees of freedom. Setting $h(p,F,T) = h_0(p,F) + a T^x$, with e.g.~$x=2$ for a metal, one obtains
at $r = 0$ a critical contribution to the specific heat
and thermal expansion, $C_{\rm cr}/T = -\partial^2_T \mathcal{V}_{\rm min} \sim |h_0|^{1/3} T^{x-2}$ and
$\alpha_{\rm cr} = \partial_T \partial_p \mathcal{V}_{\rm min} \sim T^{x-1} |h_0|^{-2/3}$, respectively, for $T \to 0$.
The critical  Gr\"uneisen ratio in this limit is
given by $\Gamma_{\rm cr} = \alpha_{\rm cr}/C_{\rm cr} =
\frac{1}{3 (x-1)}\frac{1}{V_m (p-p_c)}$ for $h_0 \propto p-p_c$ and the prefactor now depends on $x$.

{\em Influence of disorder --}
We briefly comment on the influence of disorder on EQC. Usually, one distinguishes the effect of {\it random mass} and {\it random field} disorder describing disorder configurations that, respectively, preserve or explicitly break the symmetry associated with the order parameter, see Ref.~\cite{Bruce:1980} for a review in the context of structural transitions. In case of symmetry-breaking EQC the importance of the former is decided by a modified Harris criterion. Due to the enhanced spatial correlation volume $\xi^{d_{\rm eff}}$ with $d_{\rm eff} = 6 -m$, random mass disorder is irrelevant or marginal for $m=1,2$, respectively,
as $2 \leq d_{\rm eff} \nu = (6-m)/2$. On the other hand, random field disorder, if present, is relevant and is expected to modify the criticality \cite{Imry:1984}.

{\em
Linear coupling to soft, electronic modes --}
Besides being of fundamental interest, the notion of EQC is actually relevant whenever a critical mode $\Phi$, e.g., of electronic origin, couples linearly to strain, $\mathcal{L}_{\rm int} = \gamma_1 \Phi \boldsymbol \varepsilon$. In this case, the elastic moduli obtain a strong perturbative renormalization by the critical susceptibility $\delta C \sim \gamma_1^2 \langle \Phi \Phi\rangle$ that is singular by definition and drives the crystal unstable. Upon approaching the phase transition, a crossover to elastic criticality occurs when this renormalization becomes of the same order as the elastic moduli themselves.
For classical elastic criticality, this is well-known \cite{Levayuk:1970,Villain:1970} and has been experimentally
confirmed for the classical metaelectric endpoint in KH$_{2}$PO$_{4}$ \cite{Courtens1973}.
The Mott endpoint at finite $T$ is also governed by critical elasticity \cite{Zacharias12}.
Such a crossover to elastic criticality is also expected for the quantum case.

A crossover to a symmetry-breaking EQC should occur whenever an electronic order parameter $\Phi$ breaks a point-group symmetry of the crystal. This is in particular the case for Ising-nematic ordering associated with a
Pomeranchuk instability of a Fermi surface \cite{nematic-selected-works} described by the order parameter
$\Phi = \Psi^\dagger(\partial_1^2 - \partial_2^2) \Psi$, where $\Psi^\dagger$ and $\Psi$ are fermionic creation and annihilation operators, respectively. Such a transition that breaks discrete rotation symmetry of a Fermi surface has been discussed for the ruthenates, the cuprates, and, recently, for the Fe-based superconductors \cite{Fradkin-2010}.
In a tetragonal crystal the order parameter $\Phi$ generically couples linearly to the strain
component $\varepsilon_{11}-\varepsilon_{22}$, thereby triggering a tetragonal-to-orthorhombic elastic transition.
Such a structural transition, and the associated quantum phase transition are currently being studied in systems
such as Ba(Fe$_{1-x}$Co$_x$)$_2$As$_2$ and FeSe tuned by pressure. In these systems, even if the primary order
parameter $\Phi$ is of electronic origin~\cite{Mazin2009,Qi2009,Cano2010,Fernandes2014,Davis2014},
the resulting quantum criticality is eventually governed by the long-range
shear forces of the crystal with the concomitant critical phonon thermodynamics as presented in
Fig.~\ref{fig1} \cite{Unpublished}, provided the transition itself stays continuous~\cite{first-order}.
Similarly, a crossover to a solid-solid QCEP is expected whenever the order parameter couples linearly to a strain
component that is invariant under point group operations.
Examples of such cases
include the metaelectric and metamagnetic QCEP \cite{Millis02,Zacharias13}, as well as
the Kondo volume collapse transition at $T=0$ \cite{Dzero2006/10,Hackl2008}.

\emph{Summary --}
We studied all the different universality classes of elastic transitions of a crystalline lattice at zero temperature, namely, those that break point-group symmetries spontaneously and solid-solid quantum critical endpoints.
Elastic quantum criticality is triggered whenever the
order parameter couples linearly with strain. Consequently, they are relevant for studying a wide range of correlated
electron systems such as the ruthenates, cuprates and certain Fe-based superconductors.

We acknowledge helpful discussions with A. Cano and A. Rosch. This work is supported by the DFG grant FOR 960.

\clearpage
\onecolumngrid

\section{Supplementary Information for ``Quantum critical elasticity''}

\newcommand{\ep}{\varepsilon}
\newcommand{\bq}{{\bf q}}

\section{An Example: Tetragonal to Orthorhombic Transition}

The purpose of this section is to supplement the general discussion of the main text
by examining in detail, as an illustrative example of an elastic instability,
a three dimensional lattice undergoing a tetragonal to an orthorhombic transition
(noted in the fourth row in Table I of main text). This transition
has been discussed in the context of the cuprates, bilayer ruthenates, and more
recently for certain
iron based superconductors such as Ba(Fe$_{1-x}$Co$_x$)$_2$As$_2$ and FeSe tuned by pressure.
In particular, our goal is to derive Eq.~(4) of the main text for this transition,
which is a special case of the more general Eq.~(2). Furthermore, we provide details of
the calculation of the temperature dependence of the critical tuning parameter
$r$, which governs the exponent $\theta$, and to show that
$ \delta r(T) \propto T^{5/2}$ for such an instability.

\subsection{Model \& dispersion of the critical mode}
The most general elastic free energy compatible with tetragonal symmetry is given by
\begin{align}
\label{eq:FE}
F_E &= \frac{c_{11}}{2} \left( \ep_{11}^2 + \ep_{22}^2 \right) + \frac{c_{33}}{2} \ep_{33}^2
+ \frac{c_{44}}{2} \left( \ep_{13}^2 + \ep_{23}^2 \right) + \frac{c_{66}}{2} \ep_{12}^2
+ c_{12} \ep_{11}\ep_{22} + c_{13} \left( \ep_{11} + \ep_{22} \right) \ep_{33} + \cdots
\end{align}
Here $\ep_{ij} \equiv (\partial_i u_j + \partial_j u_i)/2$, with $(i,j) = (1, 2, 3)$ are the strains, the vector
${\bf u}$ denotes displacement from equilibrium, and $c_{11}$ etc.\ denote elastic
constants in Voigt notation. The ellipsis denote terms involving higher powers of $\ep_{ij}$,
as well as those involving gradients of the strains.

The dynamical matrix $\hat{M}(\bq)$ is defined by the relation $F_E^{(2)} \equiv u^{\ast}_i(\bq) M_{ij}(\bq) u_j(\bq)/2$,
where summation over repeated indices is implied, and $F_E^{(2)}$ is the quadratic part of the free energy involving the
uniform strains. We get
$M_{11} = c_{11} q_1^2 + c_{66} q_2^2/4 + c_{44} q_3^2$,
$M_{22} = M_{11} (q_1 \leftrightarrow q_2)$,
$M_{33} = c_{33} q_3^2 + c_{44} (q_1^2 + q_2^2)/4$,
$M_{12} = (c_{12} + c_{66}/4) q_1 q_2$,
$M_{23} = (c_{13} + c_{44}/4) q_1 q_3$,
$M_{23} = M_{13} (q_1 \leftrightarrow q_2)$.
We write ${\bf u}(\bq) = \sum_{\mu} U_{\bq, \mu} {\bf n}_{\bq, \mu}$,
where ${\bf n}_{\bq, \mu}$ are the polarization vectors, $U_{\bq, \mu}$ are the associated displacements at $\bq$, and
$\mu = (\alpha, \beta, \gamma)$ is the polarization index.
The eigenvalue equation $\hat{M}(\bq) {\bf n}_{\bq, \mu} = \rho \omega_{\bq, \mu}^2
{\bf n}_{\bq, \mu}$, with the ionic mass density $\rho$, defines the (bare) phonon dispersions $\omega_{\bq, \mu}$.

In the model defined by Eq.~(\ref{eq:FE}) the critical modes (for which the velocities vanish) are restricted
on the $q_3=0$ plane, along the two lines $q_1 = \pm q_2$. Thus, the critical manifold is one-dimensional (i.e.,
$m=1$).
It is easy to diagonalize $\hat{M}$ on this manifold. For $q_1=q_2$, $q_3=0$ we obtain that
the critical polarization vector is
${\bf n}_{\alpha} = (1, -1, 0)/\sqrt{2}$ with the eigenvalue $\rho \omega_{\bq,\alpha}^2 = (c_{11} - c_{12}) q_+^2/2$,
where $q_+ \equiv (q_1 + q_2)/\sqrt{2}$. Note that, the phonon velocity vanishes at the tetragonal to orthorhombic
transition where $c_{11} - c_{12} \rightarrow 0$.
On this line the non-critical polarization vectors are
${\bf n}_{\beta} = (1, 1, 0)/\sqrt{2}$ and ${\bf n}_{\gamma} = (0, 0, 1)$ with eigenvalues
$\rho \omega_{\bq,\beta}^2 = (c_{11} + c_{12} + c_{66}/2) q_+^2$ and $\rho \omega_{\bq,\gamma}^2 = c_{44} q_+^2/4$ respectively.
Note that, for these polarizations the phonon velocities stay finite at the instability.
On the other hand, along the line  $q_1= - q_2$, $q_3=0$ we obtain that the critical polarization vector is
${\bf n}_{\beta}$ with the eigenvalue $\rho \omega_{\bq,\beta}^2 = (c_{11} - c_{12}) q_-^2/2$, where
$q_- \equiv (q_1 - q_2)/\sqrt{2}$. It is easy to check that for all other ${\bf q}$ the velocities of all the
three acoustic phonons stay finite at the transition.

One can also diagonalize $\hat{M}$ perturbatively for small deviations from the two critical directions.
Around the $q_1=q_2$, $q_3=0$ line, and for $q_+ \gg (q_-, q_3)$, the critical phonon dispersion is given by
\begin{equation}
\label{eq:critical-dispersion}
\rho \omega_{\bq,\alpha}^2  \approx \frac{c_{11} - c_{12}}{2} q_+^2 + \frac{2 c_{11} + 2 c_{12} + c_{66}}{4}  q_-^2
+ \frac{c_{44}}{4} q_3^2 + a q_+^4,
\end{equation}
where the $a q_1^4$ term, obtained from the dependence of $F_E$ on the gradient of the critical strain (not shown explicitly
in Eq.~(\ref{eq:FE})), has been added by hand. This is Eq.~(4) in the main text.
At criticality $c_{11} - c_{12}=0$, and $a q_1^4$ gives the lowest order
$q_1$-dependence of the dispersion.
The important point here is that, since the remaining  elastic constants
are finite at the tetragonal to orthorhombic instability, in this particular example $(q_-, q_3)$ are non-critical momenta,
and they scale as two powers of the critical one, i.e., $(q_-, q_3) \sim q_+^2$.
This eventually leads to an
effective spatial dimension $d_{\rm eff} = 2d - m = 5$, and the dynamical critical exponent $z=2$
(since frequency $\omega \sim q_+^2$).
Thus, Eq.~(\ref{eq:critical-dispersion}) provides a concrete example of Eq.~(2) of the main text.
A similar expression can be obtained for the dispersions around the second
critical line $q_1 = -q_2$, $q_3 =0$ for which $q_+ \leftrightarrow q_-$, and its presence simply gives a factor two to the
critical free energy.

\subsection{$T$-dependence of $r(T)$}

For tetragonal to orthorhombic transition the critical tuning parameter $r$ is the orthorhombic elastic
constant $(c_{11} - c_{12})/2$. At quantum criticality its $T$-dependence is generated by the
non-linear terms of the free energy. Denoting these contributions as $F_E^{(>2)}$, the symmetry-allowed lowest order
non-linear terms are
\begin{equation}
F_E^{(>2)} = u_3 \ep_o^2 \ep_v + u_4 \ep_o^4.
\end{equation}
Here $\ep_o \equiv \ep_{11} - \ep_{22}$ is the critical strain, $\ep_v \equiv {\rm Tr} (\hat{\ep})$ is the
striction term measuring change in the volume of the elastic system, and $(u_3, u_4)$ are coupling constants.
In the following we show that both the cubic and the quartic couplings lead to $T^{5/2}$ dependence of $r$.
For this we consider the vicinity of the critical manifold $q_z=0$, $q_1 = q_2$ (the contribution of the
second critical line simply doubles the critical contribution). Keeping track of only the critical polarization
we write ${\bf u}(\bq) = U_{\bq,\alpha} {\bf n}_{\bq, \alpha} + \cdots$, with the ellipsis denoting the non-critical
contributions. In Fourier space $\bq = (q_1, q_2, q_3)$ the strains $\ep_o$ and $\ep_v$ can be expanded as
\begin{equation}
\ep_o = q_+ U_{\bq, \alpha} + \cdots, \quad {\rm and} \quad \ep_v = q_- U_{\bq, \alpha} + \cdots .
\end{equation}
The $T$-dependence from the cubic coupling is given by (ignoring numerical pre-factors)
\begin{equation}
\delta r^{(3)} (T) = u_3^2 \langle \ep_o^2 \ep_v^2 \rangle = u_3^2 \frac{1}{\beta} \sum_{\omega_n} \int d^3 q
q_1^2 q_2^2 D_{\alpha}(\bq, i \omega_n)^2 \propto T^{(12 -m - 3z)/z}.
\end{equation}
Here $ D_{\alpha}(\bq, i \omega_n) = (\omega_n^2 + \omega_{\bq, \alpha}^2)^{-1}$ is the critical phonon propagator, with bosonic Matsubara frequency $\omega_n$,
$\omega_{\bq,\alpha}$ given by Eq.~(\ref{eq:critical-dispersion}), and $\beta$ is inverse temperature. The final
$T$-dependence is obtained simply from power counting. Similarly, the contribution of the quartic coupling
is
\begin{equation}
\delta r^{(4)} (T) = u_4 \langle \ep_o^2 \rangle = u_4 \frac{1}{\beta} \sum_{\omega_n} \int d^3 q q_1^2
D_{\alpha}(\bq, i \omega_n) \propto T^{(8 -m -z)/z}.
\end{equation}
Note that, the power counting argument and the above two equations hold as well for $m=2$ transitions.
Since, typically $z=2$, both the above contributions give $T^{(6-m)/2}$ dependence.
For gapped systems this translates into an exponent $\theta = 2/(6-m)$ which governs the finite temperature
phase boundary via $T_c (r) \sim |r|^{\theta}$.
For $m=1$ our result matches with that of A. Cano and A. P. Levanyuk in Phys. Rev. B {\bf 90},
064104 (2004) that was obtained using a different method in the context of quantum critical piezoelectric ferroelectrics.
However, they considered only the cubic coupling, and not the quartic one.


\begin{thebibliography}{}
\bibitem{Remark}
We implicitly assumed the critical phonons to be ballistic. The acoustic phonon attenuation in a metal is
weak (and even weaker in a BCS superconductor). This is in contrast to quantum criticality in
metals where the order parameter is typically overdamped.
\bibitem{Unpublished}
Details of the crossover from quantum critical electronic nematicity to critical elasticity will be presented elsewhere.
\end{thebibliography}
\end{document}